# Creation and Detection of German Voice Deepfakes

Vanessa Barnekow, Dominik Binder, Niclas Kromrey,
Pascal Munaretto, Andreas Schaad and Felix Schmieder

Offenburg University of Applied Sciences, Germany

{vbarneko,dbinder,nkromrey,pmunaret,fschmied}@stud.hs-offenburg.de

andreas.schaad@hs-offenburg.de

**Abstract**

Synthesizing voice with the help of machine learning techniques has made rapid progress over the last years [1] and first high profile fraud cases have been recently reported [2]. Given the current increase in using conferencing tools for online teaching, we question just how easy (i.e. needed data, hardware, skill set) it would be to create a convincing voice fake. We analyse how much training data a participant (e.g. a student) would actually need to fake another participants voice (e.g. a professor). We provide an analysis of the existing state of the art in creating voice deep fakes, as well as offer detailed technical guidance and evidence of just how much effort is needed to copy a voice. A user study with more than 100 participants shows how difficult it is to identify real and fake voice (on avg. only 37 percent can distinguish between real and fake voice of a professor). With a focus on German language and an online teaching environment we discuss the societal implications as well as demonstrate how to use machine learning techniques to possibly detect such fakes.

## 1 Introduction

The artificial generation of natural sounding speech was a problem for quite a long time. Only in recent years, based on advances in deep learning, it was possible to generate speech that was close to recorded human speech. This progress in text-to-speech synthesis resulted in the possibility to create new products or improve existing ones such as speech assistants, navigation systems or accessibility systems for visually handicapped individuals. But there is also a downside to this technical progress. The generation of nearly human sounding speech made it possible to synthesize fake voice of every individual as long as an attacker had enough voice material from the target to train a neural network. This can lead to criminals using synthesized voice to perform, for example, phishing attacks and fraud. The Wall Street Journal reported in 2019 that a company transferred € 220,000 to criminals after they pretended to be the parent company's senior executive [2].

The goal of this work is to determine the effort required by an attacker to create a realistic audio deepfake in a German online teaching scenario using "off-the-shelf" methods. Therefore, we consider in detail the necessary steps for the realization of this process, with a special focus realistic data acquisition, which we investigate by means of two practical case studies. In order to evaluate how realistic the created voices are, we evaluated them by over 100 participants in the form of a survey. Here we specifically focused on faking a professors voice in online teaching. Subsequently, we will address the question of whether and how fake voices can be detected and what positive and negative impacts the ability to create fake voices might have on various areas of society.

This work is split into five sections. Section 2 gives an overview of Tacotron 2, a popular neural network architecture for speech synthesis. Sections 3 and 4 show how we trained different models for the German language based on a publicly available dataset of the German chancellor Angela Merkel as well as a custom dataset and model we created of a professor on our faculty. For evaluation purposes, we conducted a survey on the second model with 102 participants. Section 5 covers possible countermeasures and detection methods. Section 6 looks at possible use and abuse cases for synthesized speech. The paper ends with a discussion of related work and some directions of future work.

## 2 Text-to-Speech Synthesis with Tacotron 2

One of the biggest breakthroughs of the past few years in the speech synthesis domain is the neural network architecture Tacotron 2 published by Google in 2018. For the first time, the generated voice was deceivingly real, almost reaching the quality of the human voice in professional recordings. This was a major improvement compared to the first version of Tacotron from 2017, which was still outperformed by classic concatenative approaches [1, p. 8]. The model architecture of Tacotron 2 consists a sequence-to-sequence feature prediction network and a modified version of WaveNet as a vocoder. To combine both networks, mel spectrograms are used as an intermediate representation for sound. The result is a framework that synthesizes speech and nearly achieves real human speech [3].

This means that we do not have an end-to-end learning process but rather have to train two independent neural networks on the same data instead. In the case of the original Tacotron 2 paper, WaveNet from Google was used as a vocoder to synthesize the samples which was already published in 2016. As we will see later on, the vocoders are usually interchangeable and we can try different combinations of vocoder networks to optimize our results, even if we use Tacotron 2. The only prerequisite is that the first neural network produces an output that is a suitable input for the second neural network. In the next section, we will focus on the Tacotron 2 model architecture because it is the most widely implemented and still keeps up with the latest model architectures when it comes to performance and audio quality. In fact, models from 2021 like

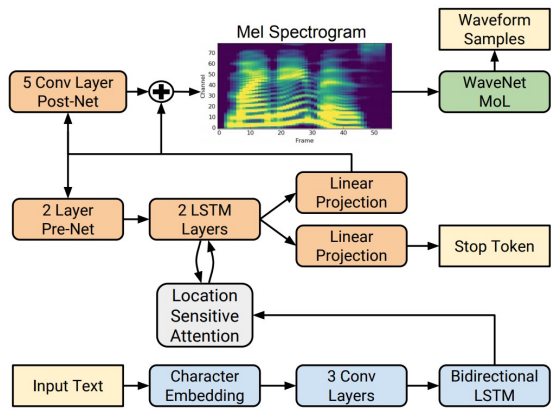

**Figure 1:** Tacotron 2 architecture [3, p. 2]

FastPitch [4] or FastSpeech [5] only achieve a Mean Opinion Score that is slightly higher than the one that was achieved with Tacotron.

## 2.1 Spectrogram Prediction Network

The Spectrogram Prediction Network consists of an encoder that converts input character sequences into a feature representation and a decoder that uses the feature representation to predict a mel spectrogram. The encoder starts by representing the input in a 512-dimensional character embedding, which is passed through three convolutional layers. These convolutional layers model long-term context in the input character sequence. The next step is to use the output of the convolutional layers and pass it through a single bi-directional long short-term memory neural network (LSTM). This step generates the encoded features the decoder needs. The encoded features are put into a location sensitive attention network. This network later summarizes each decoder output as a fixed-length context vector. The usage of an attention network encourages the model to move forward and therefore mitigates possible problems where some subsequences are repeated or ignored. The last part of the network is the decoder that predicts the mel spectrogram. The decoder starts with a small pre-net to bottleneck the incoming information. After that, the pre-net output is passed through two uni-directional LSTM layer followed by a linear transformation of the output to predict the spectrogram. As the last step in the decoder, the predicted mel spectrogram is passed through a five layer convolutional post-net. The post-net predicts a residual that is later used to optimize the predictions by minimizing the summed Mean Squared Error (MSE) [3].

## 2.2 Modified Wavenet

The modified version of WaveNet [6] is used to invert the predicted mel spectrogram into speech. The main difference between the old and the new version is the input that is required by the network. Whilst the old version needed linguistic features, predicted log fundamental frequency and phoneme durations, the new version only needs a mel spectrogram. To acquire the needed input for the old WaveNet elaborate text analysis systems as well as a robust pronunciation guide where required. These resources a lot more expensive to calculate compared to the mel spectrogram [3].

## 3 Cloning the Voice of the German Chancellor

The first goal is to create audio deepfakes of the German chancellor Angela Merkel based on a public dataset of her speeches and interviews, showing that realistic audio deepfakes can be created with reasonable effort by a technically skilled person (BA in Computer Science) that has limited computational resources and no prior knowledge in the area of speech synthesis. We will also be discussing some pitfalls and problems that may occur during the training process. Each phase of the pipeline will be explained in detail, starting from what steps we have to take to obtain an initial dataset and ending with a discussion of the audio samples that were synthesized in the course of this work using a combination of custom and pretrained models. Another goal is to find out how much training material is actually required to generate convincing deepfakes. For this, we will artificially reduce our dataset in size, train the model multiple times compare the results.

### 3.1 Data Ingestion

Training neural networks usually requires a lot of training data to achieve state-of-the-art performance. It is not uncommon that image classification architectures like DenseNet or Inception are trained on millions of samples and approximately the same applies to the speech synthesis domain. Therefore, the first task is to collect as much audio material of the person from whom we want to clone the voice as possible. Note that a synthesized voice can only be as good as the quality of the audio files it was trained on.

Since Angela Merkel is a public figure, many speeches and interviews exist on the internet which can easily be collected, for example through the official websites of the German Parliament[1], Federal Chancellery[2] or platforms like YouTube. In most cases, an official transcription is provided with the videos so the amount of manual work is kept to a minimum, especially if we are using techniques that automatically synchronize the transcript with the audio. This process, commonly referred to as forced alignment, will be used later on in Section 4 to create a custom dataset. In a nutshell, the creation of a suitable dataset is usually just a matter of days and it is definitely feasible for less sophisticated attackers, at least when it comes to public persons such as celebrities or politicians.

However, we will choose a simpler solution in our work since different prefabricated and ready-to-use datasets already exist for Angela Merkel that can be directly used as an input for Tacotron 2. One of these is the German M-AILABS dataset that was published in late 2018 for research purposes and tasks such as speech recognition and speech synthesis.[3] Besides many other German speakers, the dataset includes around 18.7 hours of transcribed voice material from Angela Merkel that is split in small files between 1 and 20 seconds. This is exactly the input format we need for our neural network architectures so we do not have to perform an extensive preprocessing. As mentioned earlier, it is also the goal to test whether acceptable results can be achieved with less training data so we create four subsets of the original dataset (see Table 1).

---

[1] https://www.bundestag.de/mediathek
[2] https://www.bundeskanzlerin.de/bkin-de
[3] https://www.caito.de/2019/01/the-m-ailabs-speech-dataset

Table 1: Train and validation sets for
the Merkel datasets with a 92/8 split ratio

| Dataset | | Train | Validation | Total |
|---|---|---|---|---|
| **Merkel Full** | Duration | 1037 min | 90 min | 1127 min |
| | # Samples | 9265 | 806 | 10071 |
| **Merkel 10h** | Duration | 570 min | 48 min | 618 min |
| | # Samples | 5060 | 440 | 5500 |
| **Merkel 5h** | Duration | 285 min | 24 min | 309 min |
| | # Samples | 2530 | 220 | 2750 |
| **Merkel 2h** | Duration | 114 min | 11 min | 125 min |
| | # Samples | 1012 | 88 | 1100 |
| **Merkel 1h** | Duration | 57 min | 4 min | 61 min |
| | # Samples | 506 | 44 | 550 |

## 3.2 Choosing an Implementation

The Tacotron 2 architecture was implemented by different people and institutions over the past few years. We did choose the implementation from NVIDIA[4] in this work, considering the fact that they modified the architecture in such a way that it produces better results without affecting the overall performance. Furthermore, it is one of the few repositories that supports Automatic Mixed Precision (AMP) which accelerates our learning process with a factor of up to three by using dedicated tensor cores on the GPU for mixed-precision computing while preserving FP32 levels of accuracy [7]. This makes it easier to train the model multiple times and test out different datasets and hyperparameters throughout the process.

If we look at the changes NVIDIA made to the original model blueprint, the first thing to notice is that only a single uni-directional LSTM layer is used in the decoder while the paper suggests two. This is a trade-off after all because on the one hand it slows down the attention learning process, but on the other hand it allows us to achieve better voice quality by further reducing the training and validation losses. Another change is the fact that dropout layers are used to regularize the LSTM layers instead of zoneout layers. In practice, it turned out that dropout layers are just as effective as their counterpart and they are considerably faster during training. However, the biggest change is the fact that the originally proposed WaveNet vocoder was replaced with WaveGlow. According to NVIDIA, this choice was made because of an improved audio quality and faster than real-time inference [8].

Note that we train our own Tacotron 2 model but for WaveGlow, we fall back to pretrained models that are provided via the NVIDIA NGC platform.[5] The reasoning behind this is the fact that it takes multiple weeks of training to converge a WaveGlow model and there is no real advantage by doing so because pretrained models generalize well to unseen speakers and languages. Nevertheless, it has to be kept in mind that WaveGlow was initially released in 2018 and other vocoders like MelGAN [9] and Parallel WaveGAN [10] exist nowadays that may outperform WaveGlow by a fair margin.

## 3.3 Data Preprocessing

The German M-AILABS dataset contains a pipe-separated metadata file with all file names of the audio clips and their respective transcriptions. This is already the correct input format for Tacotron 2 so no further changes have to be made. What has to be adjusted however is the charset in the text processing component of the Tacotron 2 implementation because characters are included in some transcriptions that are not part of the default configuration (e.g. ß, ä, ü or ö). After we adjust the charset, we remove all other symbols that are not part of it, otherwise they could cause errors during training. For this, we can define a custom text cleaner function that will be used by the model to preprocess every text sequence during training and inference. Besides the removal of unwanted characters, the function should at least perform the following text normalization tasks to make the most of the training set:

(1) Transform all characters to lowercase
(2) Write out all numbers
(3) Replace all abbreviations with the full word

A more difficult challenge for our model are words that are spelled the same but pronounced differently. In linguistics, such words are called heteronyms. If we take English word "present" as an example, the following pronunciations are possible:

- prɪˈzent as in "presenting to the class"
- prez(ə)nt as in "the period of time right now" or "gift"

During inference, it is not possible for the model to decide which pronunciation is correct without fully understanding the sentence context. However, this would exceed the scope of a text-to-speech model. To solve this problem, it is not uncommon to train the model on phonemic orthography where the written symbols correspond to the actual spoken sounds. Fortunately in our case, heteronyms rarely occur in the German language and the pronunciation is very close to the written representation. Anglicisms on the other hand will be challenging and the model will fail to pronounce English words such as "team". Instead of converting the dataset to phonemes, an easier solution in our case is to normalize the text during inference. For example if we want to pronounce the word "team" in a sentence, we can simply replace it with "tiem".

---
[4] https://github.com/NVIDIA/tacotron2

[5] https://ngc.nvidia.com

Table 2: Models trained on a NVIDIA RTX 3090

| Dataset | Batch Size | Steps | Validation Loss |
|---|---|---|---|
| Merkel Full | 56 | 140.000 | 0.1604 |
| Merkel 10h | 8 | 30.000 | 0.2685 |
| Merkel 5h | 8 | 30.000 | 0.2921 |
| Merkel 2h | 8 | 30.000 | 0.4083 |
| Merkel 1h | 8 | 30.000 | 0.5682 |

When it comes to audio preprocessing, a general advice is to trim any silence (i.e. no one speaking) at both ends of all audio files and then add 0.3 to 0.5 seconds of silence (i.e. real silence) at the end. This helps with the attention learning process, especially when we are training the model from scratch. Another thing we should take care of is to sort out all audio clips that are either less than half a second or longer than 30 to 40 seconds because they can lead to errors during training. Finally, we have to make sure that all audio files are encoded as 16-bit PCM waveforms and the sample rate is equal to 22,050 kHz. This can be done by the SoX command line utility on Linux for example. The default Tacotron 2 configuration is optimized for these audio properties and otherwise we would have to adjust different hyperparameters in the model configuration such as the filter length, hop length and window length.

### 3.4 Training

Before starting the training process, the following heuristics should be taken into account because they can greatly influence the training speed as well as the quality of our synthesized samples:

- **Warmstart:** The implementation from NVIDIA offers the possibility to drop the embedding weights of a pretrained model by passing an additional `--warm-start` parameter to the training script. This erases the learned voice but keeps the linguistic characteristics intact that can be transferred to other speakers and languages, effectively reducing the time until convergence. We will be using a pretrained Tacotron 2 model from NVIDIA as our starting point that was trained for 1,500 epochs on the English LJSpeech corpus. Normally we expect to see alignment after 400 to 500 epochs, but with the help of a warmstart, we already start to see first signs of alignment after 5 to 10 epochs.
- **Learning Rate Reduction:** The initial learning rate of 1e-3 should be reduced throughout the learning process if the training loss stagnates or loss spikes can be observed. In the latter case, we should recover the last checkpoint before the spike and continue to train with a reduced learning rate. Learning rate reduction has a noticeable effect on the audio quality, especially if we train on a reduced dataset. A more modern approach would be to modify the code and use an adaptive learning rate with exponential decay as other implementations do.
- **Batch Size:** Another hyperparameter that has great impact on the results is the batch size. Larger batch sizes allow us to process more data at once and fully utilize the GPU resources, but we do not benefit from the regularization effect of smaller batch sizes. It turned out that training with a batch size of 32 to 48 is the sweet spot if we train on the full dataset but we have to reduce it to a smaller number if we reduce the length of our dataset, otherwise the model does not learn any attention.

A summary of the training process is shown in Table 2. The reference model was trained on the full dataset for the longest and therefore it is expected to produce the best results. The training of the other models was stopped early because there was no improvement anymore after 30,000 iterations. For comparison reasons, all models except the reference one are trained on a batch size of eight.

The loss curves for the training process are shown in Figure 2. Note that even though the validation loss may increase over time, the model may still improve in terms of voice quality. This means besides the validation loss, we should also check the alignment and quality of synthesized samples throughout the training process.

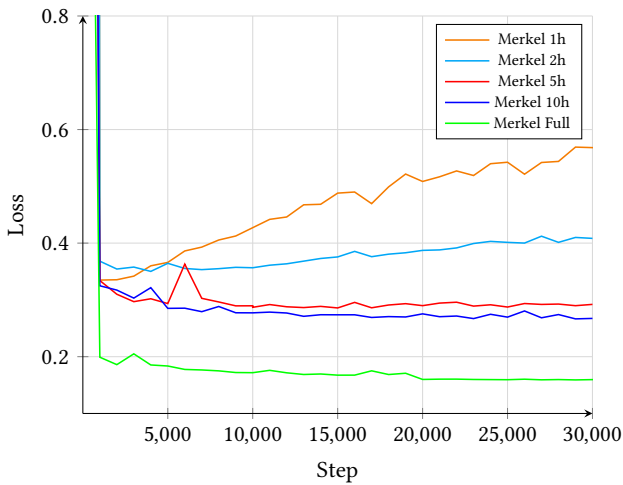

Figure 2: Validation loss curves for the first 30,000 steps of the Merkel datasets

### 3.5 Inference

After finishing the training process for all our models which takes roughly eight to nine days on a NVIDIA RTX 3090, the checkpoints can be used for inference to generate mel spectrograms based on text. For this, we can use the code that is provided with the NVIDIA implementation. The next step is to transform the mel spectrogram to speech and as already mentioned earlier, we will use a pretrained WaveGlow model from NVIDIA for this. Similar to the pretrained Tacotron 2 model we used for a warmstart, the WaveGlow model was trained on the English LJSpeech corpus for 3,000 epochs. One thing to note is that we can add an optional denoising step during the inference by using the respective code from the WaveGlow repository. This allows us to remove some of the model bias from the final sample, effectively eliminating background noises and high whistle like sounds. We can also set the strength of the denoiser and as it turned out, a value of 0.1 is the best compromise between reducing unwanted artifacts and not affecting the speech quality itself. An alternative is to manually edit the samples ourselves by using sound editing software. In rare cases it can occur that the last frames of the generated samples sound robotic. A simple solution is to add a padding word at the end that will be cut out afterwards.

## 3.6 Results & Future Improvements

In total, 23 sentences were synthesized using each of the five models to evaluate the overall speech quality. In the case of our reference model that was trained on the full dataset, the generated voice is nearly indistinguishable from the original. The model that was trained on ten hours of data still produces surprisingly good results and there is almost no difference to the reference model. The first decline of quality can be observed when we reduce the amount of training data to five hours. The model begins to have problems with the pronunciation of some words but the overall voice quality is still good if we consider the fact we cut the original dataset by almost 75%. For the two hours model, the quality more and more depends on the sentence that was synthesized and around 50% of the samples can be immediately identified as artificially generated because of artifacts and wrong pronunciations. The one hour model does not produce usable results anymore.

In a nutshell, it is difficult to evaluate whether changing a hyperparameter improves the speech quality due to two reasons. First of all, evaluating the quality of speech is subjective and to some it may sound better or there is no noticeable change at all. On the other hand, training takes multiple days on a high-end GPU even if AMP is enabled. This makes the evaluation significantly more difficult in comparison to simple machine learning classification tasks. Nevertheless, there are several things that can be done in future work to improve the training results, especially when we want to optimize the performance on smaller datasets:

- **Coldstart:** Transfer learning from a pretrained model can be useful as a proof of concept, but to achieve the best results, the model should be learned from scratch. Experience has shown that we should start to see alignment between epoch 400 and 600, however, there were no signs of alignment after more than 900 epochs which was equal to 15 days of non-stop training. As a result, the training process was stopped. The reason could be that the Merkel dataset has a worse quality than LJSpeech, as there were too many incorrect transcriptions or a batch size of 56 was too small for a cold start, considering the fact that NVIDIA trained there model on eight V100 GPUs with a batch size of 128. Another improvement would be to train the vocoder from scratch. Survey papers have shown that some vocoders generalize better to unseen speakers and languages, however, all models share the problem that there is still a large degradation of quality [11]. For example, ParallelWaveGAN achieves a Mean Opinion Score (MOS) of 4.59 for a seen speaker and language, but if we use the same model for an unseen speaker and language, the mean opinion score is only 2.17 which is quite bad. We should therefore not only consider to use other vocoders, but also train them from scratch as we did with our Tacotron 2 model.

- **Different TTS Models:** Tacotron 2 is one of many model architectures for text-to-speech. Alternatives like TransformerTTS, FastPitch, FastSpeech or GlowTTS exist that may produce better results, at least for smaller datasets.

- **Different Vocoders:** As mentioned earlier, WaveGlow may not be state of the art anymore. Alternatives like MelGAN or Parallel WaveGAN exist that achieve a higher MOS than WaveGlow, at least according to their paper. An unofficial implementation of MelGAN[6] was tested in the course of this work, however, the results were worse than WaveGlow. A reason for this could be that the repository is not optimized in the same way as WaveGlow or the model architecture simply does not generalize well to unseen speakers and languages. A solution would be to not use a pretrained MelGAN from the repository but rather train and compare our own models instead.

- **Data Augmentation:** A common technique in machine learning to enlarge a dataset is to create slightly modified copies of existing samples. For speech synthesis, this could mean to slightly change the speaking speed or pitch to get different tones in the voice or to add a white noise with zero mean and a unit variance. Another possibility is to extract segments of sentences and add them to the dataset separately. These approaches were suggested by one of the NVIDIA contributors of Tacotron 2 but did not yield any improvements when they were tested for the one hour Merkel dataset.

## 4 Cloning the Voice of a University Professor

Based on our previous results, we can expect a model to generate considerably good samples if it is trained with two to three hours of high quality audio data. Accordingly, the goal is now to confirm this observation using a custom dataset from someone "we know". For this purpose, we had three hours of audio recorded by a university professor. Half of the data originated from synchronous Zoom online lectures (with a JABRA Speak 510 as input device) and the other half was professionaly recorded in an asynchronous online lecture (with a Rhode NT1-A microphone and Arturia Audiofuse DAC9). We then used aeneas[7] to automatically synchronize the transcript with the original audio file. The timestamps are then used to extract all sentences from the recording. The results are roughly 1800 audio files with lengths between 2 to 20 seconds as well as a new metadata file with all transcriptions. The full process is summarized in Figure 3. Next, we divide the metadata file in a train and validation set as shown in Table 3 and we should also make sure not to surpass a lower boundary of 100 validation samples. Normally we would increase the ratio of the validation set (e.g. 60/40) if we are training on less data, however, it turned out that we should do exactly the opposite for speech synthesis and rather learn on as much data as possible. Table 4 and Figure 4 summarize the training process which is identical as before, the only difference being the fact that we trained the model for a longer period of time. Interesting is the fact that the validation loss is slightly better than the one that was achieved in the last section on ten hours of data. The reason for this is the fact that the overall quality of the audio files in the custom dataset is better than the Merkel dataset.

---

[6] https://github.com/seungwonpark/melgan
[7] https://github.com/readbeyond/aeneas

Table 3: Train and validation sets for the custom dataset with a 92/8 split ratio

| Dataset | | Training | Validation | Total |
|---|---|---|---|---|
| **Prof Full** | Duration | 186m | 12m | 198m |
| | # Samples | 1670 | 100 | 1770 |

Table 4: Model trained on a NVIDIA RTX 3090

| Name | Batch Size | Iterations | Validation Loss |
|---|---|---|---|
| **Prof Full** | 8 | 90.000 | 0.2494 |

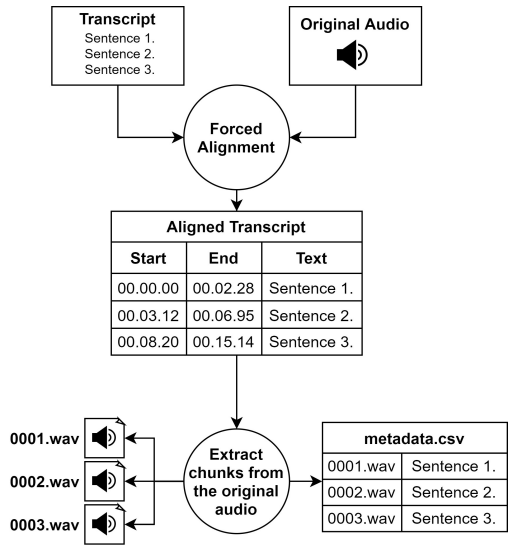

Figure 3: Forced alignment with aeneas

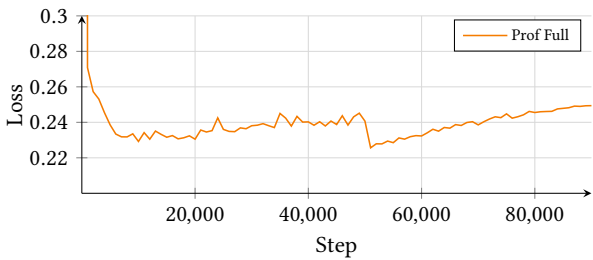

Figure 4: Validation loss curve of the custom dataset

### 4.1 Survey

In order to test how convincing our deepfakes are, we conducted an online survey with 102 participants. The initially expected goal of our study was to obtain a result showing that the fake voices cannot be correctly detected in more than 50% of the cases. The survey included ten real audio files, recorded as such by the professor with a Rhode NT1-A microphone and an Arturia Audiofuse DAC using Audacity, and eleven deepfakes, with sentences he had never said, such as "Please enter A+ as a grade" or "The government fails in controlling the pandemic situation and this is horribly to look at."

Below are six criteria to obtain the most representative survey possible in the context of this work:

(1) **Unseen Words**: The deepfake sentences contained words that did not appear in the training set.

(2) **Different Contents**: The real and fake audio files are made up of realistic and unrealistic statements, so as not to be able to make a statement about authenticity based on the content.

(3) **Different Quality**: Since the real audio files were recorded with a professional microphone, the quality had to be degraded. So we processed the audio samples with Audacity and combined random side-effects like noise, compressor, reverb, bass and treble. The reason for this is that, on the one hand, no decision should be made based on equal sounding qualities. On the other hand, many attack scenarios are based on unprofessional equipment (e.g. phone calls, mobile phone videos, voice messages, etc.) where the quality of the audio files is poor due to room acoustics, wind and other background noises or due to the limited transmission rate.

(4) **Odd Number of Audio Files**: We took 21 audio files to prevent participants from assuming an equal distribution of real and fake files when reading, for instance, 20 audio files, thus reducing an influenced decision.

(5) **Listening Once**: To represent a realistic scenario, we asked (but did not force) the subjects to listen to the audio files only once and evaluate them directly. This is to avoid comparing the audio files with each other, which would make it easier to decide on the authenticity.

(6) **Different Subjects**: In order to obtain a realistic picture of the survey, different groups of people between the age of 18 and 64 were questioned, some of whom were familiar with the professor's voice and some of whom were not. The first group consisted of 19 students who knew the professor only through the online lectures. The second group was made up of 14 students who knew the professor in person. Both groups included individuals between the ages of 18 and 31. The third group consisted of 17 faculty members and colleagues of the professor, ranging in age from 29 to 64. The last and largest group, with 52 participants, included 40% under the age of 30 and about 66% overall who did not know the professor's voice.

Within the scope of this work, the following analysis is based on basic statistical techniques and does not include other in-depth analysis techniques. During the survey, participants rated the audio files using a scale with the options "real", "rather real", "rather fake", "fake" and "no idea". Table 5 shows an overview of all results. The answers with "no idea" (14% of all answers) have not been included in the evaluation and "rather fake" and "rather real" have only been weighted half as much. For each group and for all groups in total, the number of participants, the percentage of correctly identified real and fake audio files, and the detected deepfakes are given. Additionally, a confusion matrix visualizes the number of correctly or incorrectly guessed audios based on the majority vote.

**Table 5:** Survey results of all groups (F = fake and R = real)

|  |  | G1 |  | G2 |  | G3 |  | G4 |  | Total |  |
|---|---|---|---|---|---|---|---|---|---|---|---|
| Participants |  | 19 |  | 14 |  | 17 |  | 52 |  | 102 |  |
| Correctly identified |  | 57% |  | 67% |  | 38% |  | 33% |  | 43% |  |
| Deepfake detected |  | 55% |  | 63% |  | 27% |  | 27% |  | 37% |  |
| Confusion matrix |  | Actual label |  |  |  |  |  |  |  |  |  |
|  |  | F | R | F | R | F | R | F | R | F | R |
| Guessed label | F | 6 | 5 | 7 | 4 | 3 | 8 | 3 | 8 | 4 | 7 |
|  | R | 4 | 6 | 3 | 7 | 5 | 5 | 6 | 4 | 6 | 4 |

The results of the survey are surprisingly good, considering that the model was only re-trained with 3 hours of audio. On average, a deepfake was only correctly detected 37% of the time, exceeding our initial hypothesis of 50% (fair coin toss). It is noticeable that the first two groups have a higher identification rate than the others. This could be because they are all students who know the professor's voice very well and were more motivated to achieve a good result. Based on their age, it can be expected that they have average better hearing than group 3 and 4. In addition, they are active in the IT field, so they have a better feeling about this topic. Furthermore, it can be assumed that the question instructions on single listening and thus on comparing the audio material were partly ignored.

Some participants gave us feedback that they decided on the basis of quality (noise, reverb, tinny sound, etc.), as they could not detect any differences in prosody. It is also important to keep in mind that participants are aware of the presence of deepfakes. In real life, people do not think about the possibility of fake material, also time pressure or other social engineering techniques are often applied to the victim, which makes it more difficult for humans to detect.

## 5 Detection

Since our previous results have shown that realistic voices can be cloned with existing methods and already comparatively small data sets of a few hours, the motivation to find adequate detection measures is accordingly high. In this section, we examine the audio data we have created to determine if and how technical measures can be applied to automatically detect synthetically generated voice. We will demonstrate how correlation-based properties of audio tracks can be used to create speech profiles, and how these profiles can be used to identify the differences between the real speaker and the synthesized voice. Subsequently, we will perform cluster-based anomaly detection on the basis of these profiles to evaluate whether this method can be used for the detection of synthesized voices.

### 5.1 Bispectral Analysis

The bispectrum is a higher order time series analysis technique that can be applied to a single time series. For a triplet of frequencies, it measures the reversibility in time and the symmetry about the mean of its flux distribution. The bispectrum is calculated as a complex number and consists of a magnitude and a phase (the biphase) [12].

In our case, the single time series of interest is the audio signal of individual samples, which we inspect using bispectral analysis. The application of this method to distinguish real and synthetic voices was demonstrated in [13]. The authors compared the voices of common voice assistants with those of real people and custom wavenet models and concluded that this method is highly effective. We adopt relevant parts of their methodology to check whether this represents a suitable measure to detect the synthetic voices created in our work.

To implement the bispectrum analysis using the python *stingray* library. This implementation allows us to calculate the biphase and magnitude of selected audio signals extracted from the WAV-files, which we will use for the following analysis steps. For the purpose of the demonstrations contained in this section, we use a subsample of the data we generated as well as a subsample of the original data. These subsamples contain respectively 100 randomly selected real and synthetic samples of the chancellor dataset and all samples of the professor dataset that were used for the survey (Section 4.1).

For a visual examination of the real and synthetic records of both subjects/persons, the magnitude and the biphase of individual audio samples are shown in Figure 5. The two axes of the subfigures represent the two frequencies that are used for the bispectral analysis. The third frequency, which is used for the bispectral analysis, is composed of these two frequencies and is therefore not shown. The coloring of the individual points represents the value of the magnitude and the phase and thus allows an interpretation of the correlation under consideration of the used frequencies. For all samples we see a high magnitude at the center of the plot, which shows an expected strong correlation of individual data points with the nearby data points. It is interesting to note that in the synthesized data, the horizontal and vertical patterns, which originate from the center, are more pronounced. When interpreting the biphase, it can be seen that the synthesized samples tend to have a stronger artifact formation.

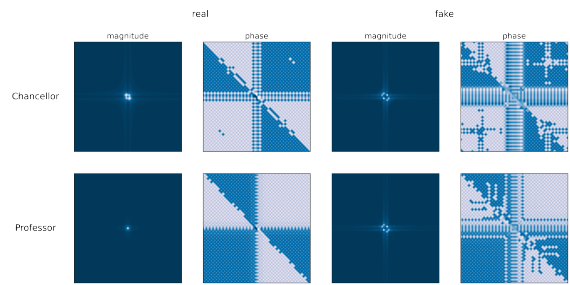

**Figure 5:** Magnitude and phase of individual samples based on the bispectral analysis

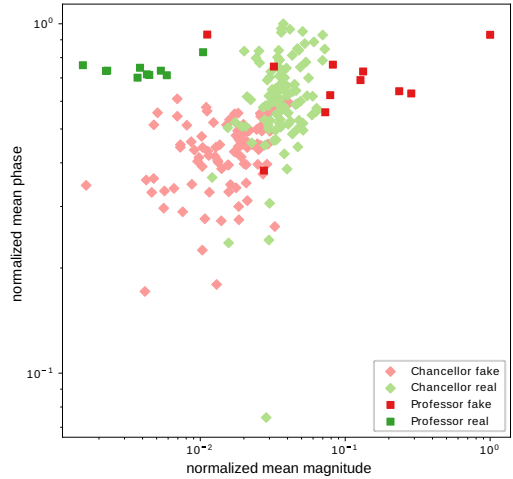

Figure 6: Fake and real voice profiles of different speakers based on bispectral features

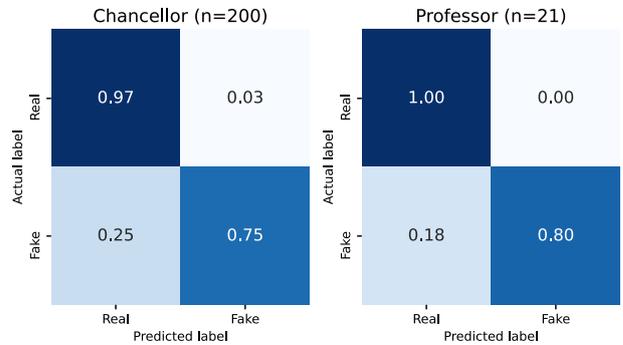

Figure 7: Confusion matrix of the cluster-based detection

In order to validate whether the magnitude and phase can be used to distinguish between real and fake data, we visualize them in a scatter plot for all 221 samples considered. To be able to represent these two-dimensionally, we use the normalized mean values of the magnitude and phase, which can be seen in Figure 6. Due to individual outliers, the graph was displayed on a logarithmic scale. Considering the subjects (chancellor and professor) individually, we see that the real and fake votes can be separated approximately linearly, although there are still clear overlaps in the center region of the graph. Comparing the samples of both groups shows that the fake samples of the professor lie closely to the original samples of the chancellor. From this observation we can conclude that the application of a countermeasure to a single person is more reasonable than trying to apply it in a generalized way. Since we only considered the mean values of magnitude and phase in the visual observation, a large amount of information is lost, which is why we do not want to base the detection of fake samples on this information alone, but want to perform a more precise detection based on an extended feature set in the following section.

## 5.2 Anomaly Detection

To generate appropriate features for the detection of the deepfakes, we adopt the methodology shown in [13], which includes the mean, variance, skewness and kurtosis for both the magnitude and the phase of the bispectral analysis. Instead of a supervised algorithm, we deliberately use a clustering procedure to detect synthesized voices as anomalies, based on the standardized features. We base this decision on the previous observation that strong deviations of real and fake voices are recognizable for different voice profiles. Our procedure is therefore based on the assumption of having audio samples of a target's real voice and matching them with a potential fake. We use the DBSCAN clustering algorithm and match the obtained clusters with the known real voice data. The results of this procedure can be seen in Figure 7. We are thereby particularly interested in the lower row of the confusion matrix, which shows how well the fakes can be detected, which in the cases shown corresponds to a precision of 75-80 %.

## 6 Use and Abuse Cases

The applications of audio deepfakes are very diverse. Some provide positive individual or societal value, while others can cause harm to individuals or even an entire society. In the following, possible application scenarios are explained and categorized according to their impact. All scenarios refer only to the use of pure audio.

### 6.1 Use Cases

Deepfakes can create possibilities and opportunities for several types of communication. They can be used to entertain or educate people and also find application in the health care sector.

- **Educational Purposes**: The educational sector offers a wide range of applications for deepfakes [14]. When using pure audio, historical figures can be imitated and thus presented close to reality. An audio-based example would be the setting to music of speeches or discussions of historical figures, which are only available in text form.

- **Entertainment and Art**: In the artistic field, there are also many possible applications for audio deepfakes. Artists who have already passed away can perform again and present new content. In the visual field, actors who have already died have made appearances in films[15]. Transferred to audio, deceased artists could record or perform new songs. It is also conceivable that in the future there will be licenses from commercial voice actors, audio book speakers or voice actors for films and series that can be purchased, thus saving effort, time and money. As technology advances, it may be possible to convert the voices of well-known personalities, such as actors in movies, into other languages, making movies available worldwide in their original sound. Furthermore, deepfakes can be used for satire or parodies of public figures.

- **Health and Social Care**: Audio deepfakes are also already being used in medicine and health services. They offer people

who are unable to speak due to physical limitations such as ALS the possibility to communicate with their own voice, provided that an appropriate amount of audio material has been recorded before the loss of the voice [15]. In other cases, the voice of a stranger, which feels "natural" to the affected person and represents their personality, is used instead [8]. Also it can help mourners to deal with the loss of loved ones by developing a digital version of them.

## 6.2 Abuse Cases

Despite the undeniable positive opportunities created by deepfake technology, it also lays the foundation for a wide variety of abuse cases. In the following section these abuse cases are divided into attacks that target an individual or an organization and attacks that can have a negative impact on a society as a whole.

### 6.2.1 Harm to Individuals or Organizations.

Attacks that fall under this category target individuals or organizations. The goal of these attacks is either to damage the target or extract benefits like money or information. The damage caused by these techniques can be severe for the target and its environment, but in most cases it has no significant impact on the society.

- **Social Engineering Attacks**: Social engineering refers to the manipulation of individuals in order to persuade them to perform specific tasks or to give away information that can be of use for an attacker [16]. The actual goals can be very diverse. It can be about extracting sensitive information such as passwords or user data. It can also be an instruction to transfer money to other people's accounts, so-called CEO fraud. All these attacks can be carried out on the basis of audio deepfakes by imitating the voice of the person authorized to make the request. In the context of universities, one example is the requesting or changing of grades by forging the voice of a professor in charge (as we demonstrated). In practice, the exact use of the audio files can be adapted to the respective target and the technical possibilities. Non-interactive methods of communication offer an advantage for the attacker, as he does not have to react to possible counter-questions or unexpected behavior on the part of his opponent. The required audio examples can be recorded in advance and played back at the required time. This includes pre-recorded voice messages or leaving a message on an answering machine. But live conversations can also be carried out with a collection of pre-recorded parts. The success of this depends to a large extent on the person on the other end of the line and possible questions asked by the target. Imitating a choleric supervisor for example, may lead to less skepticism in the case of requests that seem urgent, since questioning instructions often seems unusual here.

- **Discrediting**: Deepfakes also represent a very effective way of discrediting individuals. In this case, the voice of a target is imitated in order to have them make statements that morally expose them or are to be considered critical. Enormous damage can be done in both the private and public spheres. If a "normal" citizen is the target of such an attack, this could lead to a loss of social acceptance, problems within the job or even disputes with friends and family. In a university context, for example, a professor might talk negatively about his colleagues or spread conspiracy theories. If the target of such an attack is a public figure, such as a politician, the consequences could, in the worst case, affect society as a whole [17]. This case is discussed in more detail under the point "Politics" found below.

- **Blackmail**: Deepfakes can also be applied in the area of blackmailing. In this case, files are created that could be used for the discrediting method explained above. However, these files are not published, but the target is forced to pay a sum of money or provide information. The fear of publication and thus the effectiveness depends on the target and the possible consequences of publication.

### 6.2.2 Harm to Society.

Attacks under this category are defined by the impact they have on society, either immediately or in the long run. Even though the intention of an attack might not be to impact society, attacks on individuals or organizations with a high public importance like politicians or courts can cause such effects. The political status also has an influence on these effects, as tense situations might lead to overreactions.

- **Judicial Systems**: Deepfakes could be used to create fake evidence in court cases in order to influence a case. This form of evidence tampering is a major threat to the judicial system according to Pfefferkorn [18]. This might lead to unrightful convictions in the worst case. But in any case it will be necessary to invest time and money to authenticate evidence if those wrongful convictions want to be avoided or at least minimized. In 2020 a mother used audio deepfakes of her husband in a child custody case in order to prove the father threatening her. In this case the audio file was forensically examined and detected as fake evidence [19]. The usage of deepfakes in court cases therefore poses threats to individuals due to false convictions, but also to the judicial system as a whole, as fighting misinformation could cost resources and if not done properly, could lead to a loss of trust towards the courts.

- **Politics**: Another potential threat created by deepfakes is disinformation within politics. Deepfakes can easily be distributed to a wide audience making use of social media or online forums [20]. A possible attack might be targeting a political party or candidate for example during an election period or in a moment of high tension within a country. "This scenario would have an immediate impact on a society, with the potential to influence the outcome of an election, to spark violence, to inspire protests or even to encourage a coup d'état" [21]. Especially in high pressure situations, the damage created by the publishing of such a video might already be done before the correctness could be verified. In situations where the trust in a government or the media is already damaged, denials of correctness might not be trusted by the population.

---

[8] https://vocalid.ai/

- **Attack on Minorities or Vulnerable Groups**: The publication of a deepfake targeting vulnerable groups or minorities could lead to social tensions continuing to rise and discrimination and hatred against these groups increasing. Examples of this behavior could be observed, for example, in the context of the "refugee crisis" in Europe in 2014 [21]. The disinformation at that time warn rarely be based on deepfakes, however, the credibility of misinformation is increased by substantiating it with audio recordings. Targeted disinformation through deepfakes could have an immediate impact on a society and by creating strong emotions, trigger protests or conflicts between different groups. In this case, even the exposure of a deepfake could have serious consequences, as the affected social groups could consider it a targeted attack.

- **Geopolitical Attack**: A state or non-state actor could publish and spread a deepfake targeting a (foreign) government to influence a geopolitical situation. For example, a manipulated audio track could show a president ordering troops to prepare to invade a disputed territory, which would have an immediate impact on international politics. This could result in the affected country responding, even if the content was not or could not be verified as fake due to lack of the necessary tools or time to make such a determination [21]. In August 2020 an explosion in Lebanon killed 154 people. It was caused by a large amount of ammonium nitrate in a storage facility. After the explosion happened, a video that was later debunked as fake was published, in which the building was hit by a missile [22]. Even though the video was detected as fake by many experts shortly after the initial release, the president of Lebanon suggested at one point that the explosion could have been caused by just such an attack [21].

In the long term, a heavy accumulation of the problems explained above could lead to a so called *disbelief by default*, which could have a disastrous effect on society. This term describes a state, in which information is no longer trusted and credibility is always at doubt. If disbelief by default becomes a common reaction to presented information, democratic discourse loses its foundation in generally accepted facts. This includes the increasing ability of people to simply dismiss evidence as manipulated or wrong [21]. First signs of this disbelief by default can already be seen during the current coronavirus crisis. Scientific facts are often ignored and the sources are marked as untrustworthy, because the results do not match with the (political) views of other individuals [23].

## 7 Conclusion

In this work, we investigated the feasibility of generating German fake voices using established methods. Modern text-to-speech models are openly accessible and can be used relatively comfortably with a basic understanding of the technology. With regard to data acquisition, it became apparent that audio data from public figures were easy to obtain. For the German chancellor scenario we considered, transcribed data sets of 18.7 hours were also freely available. Even if an own dataset has to be created, the effort would be manageable and a matter of days, considering that speeches, interviews and transcriptions are available on the internet. Using this dataset, we re-trained Tacotron 2 based on an English pre-trained text-to-speech model from NVIDIA and obtained comparatively good German voice deepfakes. Even though a high-end GPU was used, the same results can be obtained with less powerful hardware.

For a more realistic attack scenario, a custom dataset of our university professor was created in the next step that only consisted of three hours of audio material. Despite the reduced amount of training data, the model was able to generate realistic samples. In a survey we conducted with 102 participants, only 37% were able to detect the artificially generated voices on average.

In comparison, technical detection based on the features of bispectral analysis provided significantly better results, as fakes could be detected with a precision of up to 80%. We consider this approach to be particularly practical because it requires only a few samples (21 for the professor's example), unlike supervised methods that require large amounts of data. Since each sample is only a few seconds long, this detection method can thus be applied individually to any attack scenario, even if only a few minutes of the audio material to be examined are available.

The use and abuse cases examined show not only the added value that deepfake technologies can undoubtedly offer, but also the associated dangers. The misuse of this technology can cause severe damage to individuals, organizations or even society as a whole. To prevent this abuse, or even the worst case scenario of a general disbelief by default, we will need reliable detection methods in the future.